\documentclass[prb]{revtex4}

\usepackage{graphicx}
\usepackage{epstopdf}%
\usepackage{amsmath,amssymb,dsfont,upgreek}

\usepackage[colorlinks=true,  linkcolor=blue, citecolor=blue, urlcolor=blue]{hyperref}

\usepackage{natbib}   

\begin{document}
\date{June 10, 2008}

\title{Renormalization group study of intervalley scattering and valley splitting in a two-valley system}

\author{Alexander Punnoose}
\affiliation{Physics Department, City College of the City University of New York, New York, NY 10031, USA}
\email{punnoose@sci.ccny.cuny.edu} 

\begin{abstract}
Renormalization group equations are derived for the case when both valley splitting and intervalley scattering are present in a two-valley system.  A third scaling parameter is shown to be relevant  when the two bands are split but otherwise distinct. The existence of this parameter changes the quantitative behavior at finite temperatures, but the qualitative conclusions of the two-parameter theory are shown to be unaffected for realistic choice of parameters.   
\end{abstract}


\maketitle

\section{Introduction}
Renormalization group (RG) studies of multi-valley two-dimensional electron gas (2DEG) systems has been very successful in quantitatively describing the transport properties of electrons confined in silicon inversion layers (MOSFETs)\;\cite{punnoose02, punnoose05, punnoose07}. In a disordered medium, for temperatures $k_BT< \hbar/\tau$, where $1/\tau$ is the elastic scattering rate, the propagating modes are diffusive, and it is now well understood that these modes play a central role in determining the transport properties at low temperatures\;\cite{aabook}. In two dimensions, in particular, the effects of diffusion are profound. The  electron-electron (\textit{e-e}) scattering amplitudes, for example, develop non-analytic corrections that result in  enhanced correlations at low energies\;\cite{sasha83}.  It has been shown that RG theory applied to a weakly disordered system is able to capture this scale (energy or temperature) dependence to all orders in the \textit{e-e} scattering amplitudes, making it the most promising analytical technique available to understand the physics of disordered systems. (Pedagogical reviews of  the RG theory can be found in Ref.\;\cite{yellowbook,pedagogical}.)

Weak disorder implies that  $\hbar/\tau< E_F$, where $E_F$ is the Fermi-energy. Typical high mobility two-dimensional semiconducting devices  have very small Fermi-energies with a scattering rate which is even smaller due to the very high mobility of the samples making it very difficult to access the diffusive region at experimentally reasonable temperatures. Si-MOSFETs on the other hand have only moderately high mobilities so that $\hbar/\tau$ is of the order of a kelvin while $E_F$ is of the order of a few kelvin. The impurity scattering in these inversion layers is short-ranged in character making quantum scattering the dominant scattering mechanism, while semi-classical effects arising from the impurity potential landscape are negligible at low temperatures. For these reasons, as noted in the beginning,  RG theory has been particularly successful in describing the properties of electrons in silicon inversion layers. (See Ref.\;\cite{punnoose07} for how the diffusive regime is identified experimentally and for a quantitative comparison of theory with experiment.)

The conduction band of an $n$-(001) silicon inversion layer has two almost degenerate valleys located close to the $X$-points in the Brillouin zone\;\cite{ando}. The abrupt change in the potential at the interface, which breaks the symmetry in the $z$-direction perpendicular to the 2D plane, leads to the splitting of the two valleys. Although intervalley scattering originates from both impurity scattering and  scattering due to \textit{e-e} interactions, the imperfections at the interface, which are distributed on the atomic scale, are the main source of the large momentum transfer $Q_0$ in the $z$-direction needed for intervalley scattering.

The RG theory developed in Ref.\;\cite{punnoose02} considered the valley degrees of freedom to be degenerate and distinct,  hence quantitative comparisons with experiments performed in Ref.\;\cite{punnoose07} were  limited to temperatures larger than the valley splitting, $\Delta_v$, and the intervalley scattering rate, $\hbar/\tau_\perp$, both scales being sample dependent. This paper develops the relevant scaling equations in the presence of   valley splitting and intervalley scattering.  

The scaling equations are presented in three different temperature regimes: (i)\;high temperature region, $T\agt T_v$ and $T_*$, where $k_BT_v=\Delta_v$ and $k_BT_*=\hbar/\tau_\perp$, (ii)\;low temperature region, $T\alt T_*$, (iii)\;and intermediate temperature region, $T_*\alt T\alt T_v$. The last of the three regions is relevant when the band splitting is large so that effective mixing of the valleys due to impurity scattering occurs only at sufficiently low temperatures; it is shown that the standard two-parameter description has to be modified in this case to include a third scaling variable which has quantitative effects at finite temperature but does not affect the asymptotic conclusions of the two-parameter theory.

\section{Diffusion modes and Fermi-liquid amplitudes} 
Electrons in valleys can be conveniently labeled using  additional valley indices $\uptau_z=\pm$. (For our purpose, the number of valleys $n_v=2$ located at $\pm Q_0 \hat{z}$, where $Q_0\approx 0.85\times(2\pi/a)$ with $a$ being the lattice constant of silicon.) This increases the number of single particle states to  (spin)$\times$(valley)=4. Since the diffusion modes, responsible for the relaxation of density and spin perturbations (and valley in our case) in a disordered system at long times and distances, formally occur via particle-hole excitations,  the corresponding number of (particle)$\otimes$(hole) diffusion modes equals 16.  This is a four fold increase from the case of one valley and has significant quantitative effects on transport as shown in Ref.\;\cite{punnoose02}. At low temperatures, some of these modes develop gaps (cut-offs) proportional to  $\Delta_v$ and $\Delta_*$ and are therefore ineffective (non-singular) for $T$ below the characteristic temperature scales $T_v$ and $T_*$\;\cite{fukuyama1,fukuyama2}, leading to quantitatively different scaling  as the temperature is varied.

\subsection{Single particle properties}

\begin{figure}[htb]
\includegraphics[width=0.5\linewidth]{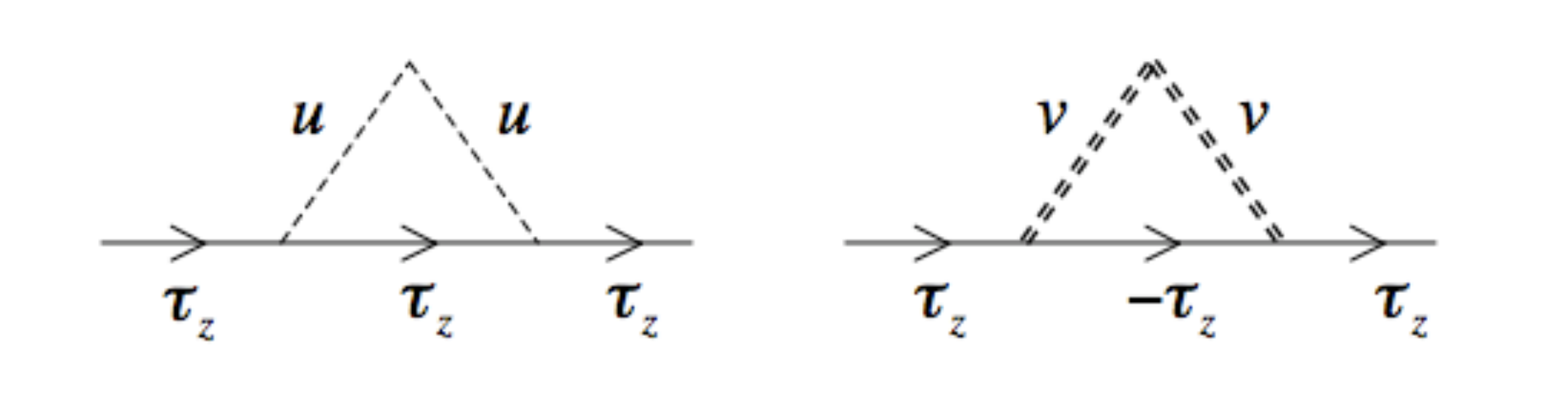}
\caption{Diagrams contributing to $1/\tau$ in Eq.\;(\ref{eqn:tau}) are shown. The properties of the intra- and intervalley impurity scatterings are defined in Eq.\;(\ref{eqn:uv}), represented here by single and  double dashed lines, respectively.}.  \label{fig:tau}
\end{figure}

At low electron densities the mobility of a 2DEG is determined by the charged centers within the SiO$_2$ layer. Due to the short-ranged nature of the impurity scattering in silicon inversion layer structures, the Drude relation for the mobility, $\mu=e\tau/m$, gives a direct measure of the single particle life-time, $\tau$. Here, $e$ and $m$ are the charge and the effective mass of the electron, respectively. Ando\;\cite{ando_valleysplitting} argued that the mobility is also determined partially by the intervalley scattering rate. To this end, the two different scattering rates, that is, the intravalley and intervalley rates, can be incorporated by introducing two scattering potentials\;\cite{fukuyama2}, $u(\mathbf{q})$ and $v(\mathbf{q})$, respectively. The potential  $u(\mathbf{q})$ is slowly varying on the scale of $1/a$ if the impurities in the oxide layer is uniformly distributed, while $v(\mathbf{q})$ is a rapidly oscillating function with momentum of the order of $1/a$. Hence the random average of the potentials $\langle u(\mathbf{q})v(\mathbf{q})\rangle=0$, with $u(\mathbf{q})$ and $v(\mathbf{q})$ satisfying
\begin{subequations}\label{eqn:uv}
\begin{eqnarray}
\langle u(\mathbf{q})u(\mathbf{q}')\rangle&=&\delta_{\mathbf{q}+\mathbf{q}'}\frac{1}{2\pi\nu\tau_\parallel}\label{eqn:u}\\
\langle v(\mathbf{q})v(\mathbf{q}')\rangle&=&\delta_{\mathbf{q}+\mathbf{q}'}\frac{1}{2\pi\nu\tau_\perp}\label{eqn:v}
\end{eqnarray}
\end{subequations}
where $\nu=m/2\pi$ is the density of states per spin and valley. The total life time, $\tau$, then equals (see  Fig.\;\ref{fig:tau})
\begin{equation}
\frac{1}{\tau}=\frac{1}{\tau_\parallel}+\frac{1}{\tau_\perp}\label{eqn:tau}
\end{equation}

\subsection{Particle-hole diffusion propagators}

The form of the particle-hole propagators (diffusons) for the impurity model defined in Eq.~(\ref{eqn:uv}) have been calculated in Ref.\;\cite{fukuyama1,fukuyama2}. The calculations are extended here to include valley splitting. 

The fluctuations in the diffuson channel, $\mathcal{D}(q,\omega)$, have  a diffusive singularity $\mathcal{D}(q,\omega)=1/(D_0q^2+|\omega|)$.  Finite valley splitting and intervalley scattering introduces gaps in $\mathcal{D}(q,\omega)$ thus cutting off the singularity.  The different diffuson modes involving fluctuations in the valley occupations are shown in Fig.\;\ref{fig:diffuson}. The details of their derivation are given in Appendix\;\ref{app:diffuson}. 

\begin{figure}[htb]
\includegraphics[width=0.5\linewidth]{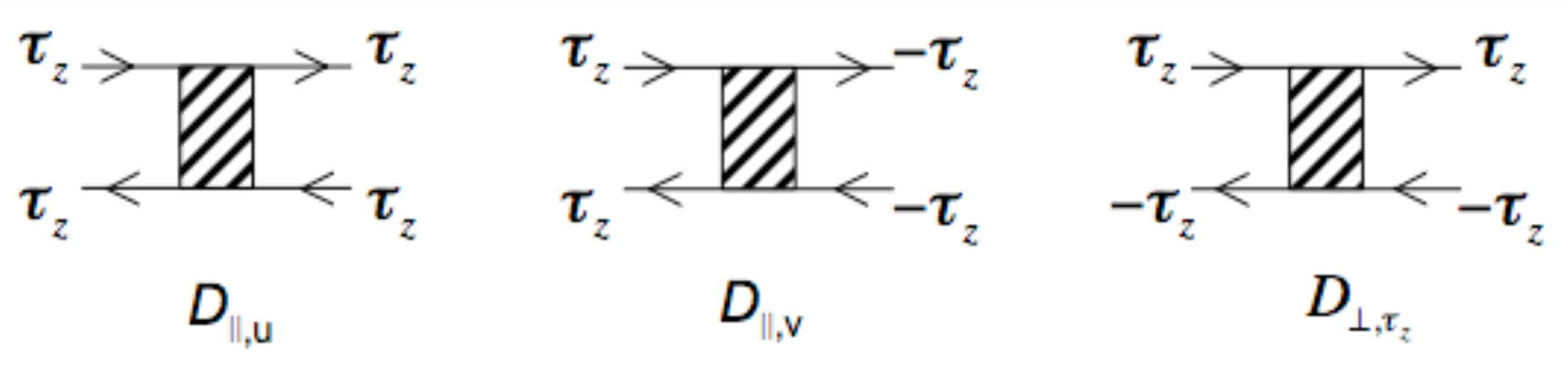}
\caption{The various diffuson blocks in the presence of valley splitting and intervalley scattering are shown.}  \label{fig:diffuson}
\end{figure}

We start by defining the elementary diffuson blocks, $\mathcal{D}_{\parallel,u}, \mathcal{D}_{\parallel,v}$ and $\mathcal{D}_{\perp,\uptau_z}$ shown in Fig.\;\ref{fig:diffuson}. The diffuson blocks $\mathcal{D}_\parallel$ are insensitive to valley splitting since both the particle and the hole (corresponding to the top and bottom lines with arrows moving to the right and left, respectively) belong to the same valley. The $\uptau_z$ index for the $\mathcal{D}_{\perp,\uptau_z}$ diffuson indicates the valley index of the particle, with the hole being in the $-\uptau_z$ valley; the two valleys are nonequivalent for finite $\Delta_v$.  
In Appendix\;\ref{app:diffuson}, the equations satisfied by the diffuson propagators are  solved in the limit of weak splitting $\Delta_v\tau \alt 1$. The solutions are expressed in terms of the diffusons $\mathcal{D}_\pm=\mathcal{D}_{\parallel,u}\pm\mathcal{D}_{\parallel,v}$ and $\mathcal{D}_{\perp,\uptau_z}$, with the corresponding gaps $\Delta_-=2\tau_\parallel/\tau(\tau_\perp-\tau_\parallel)$ and $\Delta_\perp=\tau_\parallel/\tau\tau_\perp$. (Note that $\mathcal{D}_+$, corresponding to the valley ``singlet" mode, is gapless, hence $\Delta_+=0$.)

In the limit when the intervalley scattering is much weaker than the intravalley scattering, i.e., $\tau_\perp\gg \tau_\parallel$, the scattering time $\tau\approx \tau_\parallel$. The gaps in this limit correspond to $\Delta_-\approx 2/\tau_\perp$ and $\Delta_\perp\approx 1/\tau_\perp$. In this weak scattering limit, relevant to high-mobility MOSFETs, the form of the diffusons obtained in Eqs.\;(\ref{eqn:app_dpm}) and (\ref{eqn:app_dperptau}) reduce to: (the overall factor $1/2\pi\nu\tau^2$ is suppressed)
\begin{subequations}\label{eqn:diffusons}
\begin{eqnarray}
\mathcal{D}_+(q,\omega)&=&\frac{1}{D_0q^2+|\omega|}\label{eqn:dp}\\
\mathcal{D}_-(q,\omega)&=&\frac{1}{D_0q^2+|\omega|+2\Delta_*}\label{eqn:dm}\\
\mathcal{D}_{\perp,\uptau_z}(q,\omega)&=&\frac{1}{D_0q^2+|\omega|-i\uptau_z\Delta_v+\Delta_*}\label{eqn:dperp}
\end{eqnarray}
\end{subequations}
where $\Delta_*=1/\tau_\perp$. The number of modes  that are effectively gapless depends on the relative magnitude of $T$ (or frequency) with respect  to the corresponding temperature scales $T_v$ and $T_*$. At high-$T$ all modes are gapless, while at the lowest $T$ only $\mathcal{D}_+$ remains gapless. 

\subsection{Electron-electron scattering amplitudes}

\begin{figure}[htb]
\includegraphics[width=0.5\linewidth]{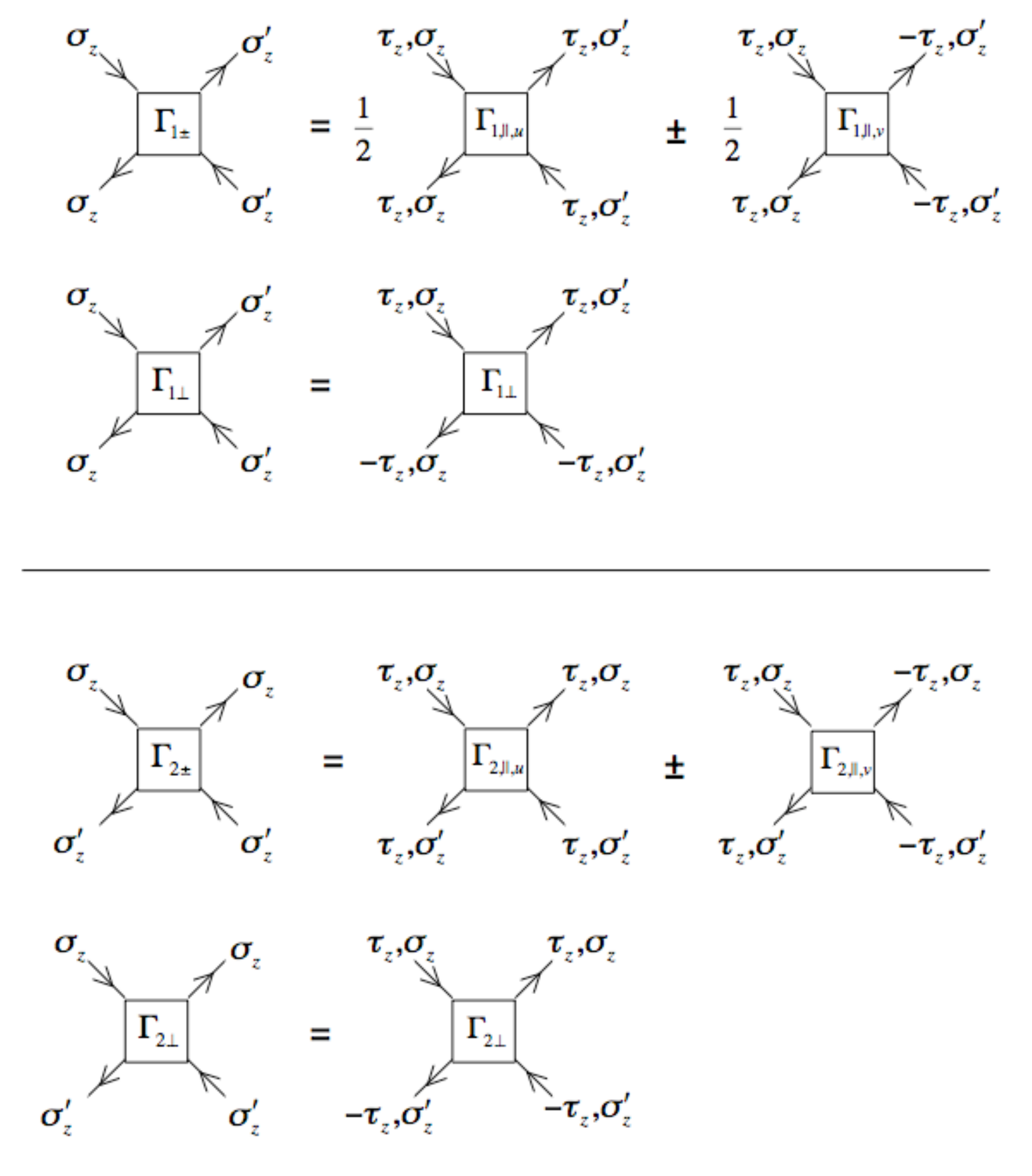}
\caption{The scattering amplitudes including the valley degrees of freedom are classified in terms of the standard static Fermi-liquid amplitudes $\Gamma_1$ and $\Gamma_2$. The same subscript convention  used to classify the diffusons in Fig.\;\ref{fig:diffuson} are used here.}  \label{fig:gamma}
\end{figure}

In this section the relevant \textit{e-e} interaction scattering amplitudes are identified.
These amplitudes are conventionally described by the standard static Fermi-liquid amplitudes $\Gamma_1$ and $\Gamma_2$ defined in terms of the spin texture of the scattering of the particle-hole pairs. The amplitudes are easily generalized  to include the valley degrees of freedom. They are shown in Fig.\;\ref{fig:gamma}. Note that the intervalley scattering amplitudes  $\Gamma_{1\perp}$ and $\Gamma_{2,\parallel,v}$ are generally negligibly small in a clean system because the Coulomb scattering involving large momentum   $Q_0$ in the $z$-direction is suppressed when the width of the inversion layer is many times larger than the lattice spacing. It is more convenient to work in the same basis as that used for the diffusons, i.e., $\Gamma_{1\pm}=\frac{1}{2}\left(\Gamma_{1,\parallel,u}\pm \Gamma_{1,\parallel,v}\right)$ and $\Gamma_{2\pm}=\left(\Gamma_{2,\parallel,u}\pm \Gamma_{2,\parallel,v}\right)$, as it allows for the amplitudes to be easily combined with the diffusion modes.

\section{Diffusion corrections}\label{sec:corrections}

It is now well understood that while the diffusion propagators when combined with \textit{e-e} scattering lead to the appearance of logarithmic corrections to the resistivity (Altshuler-Aronov corrections),  the \textit{e-e} scattering amplitudes themselves develop logarithmic corrections due to the slow diffusive relaxation\;\cite{sasha83}.  In this section, these logarithmic corrections  are obtained self-consistently in the limit of  weak valley splitting ($\Delta_v\tau \alt 1$) and weak intervalley scattering ($\tau_\perp\gg \tau_\parallel$).


\begin{figure}[htb]
\includegraphics[width=0.5\linewidth]{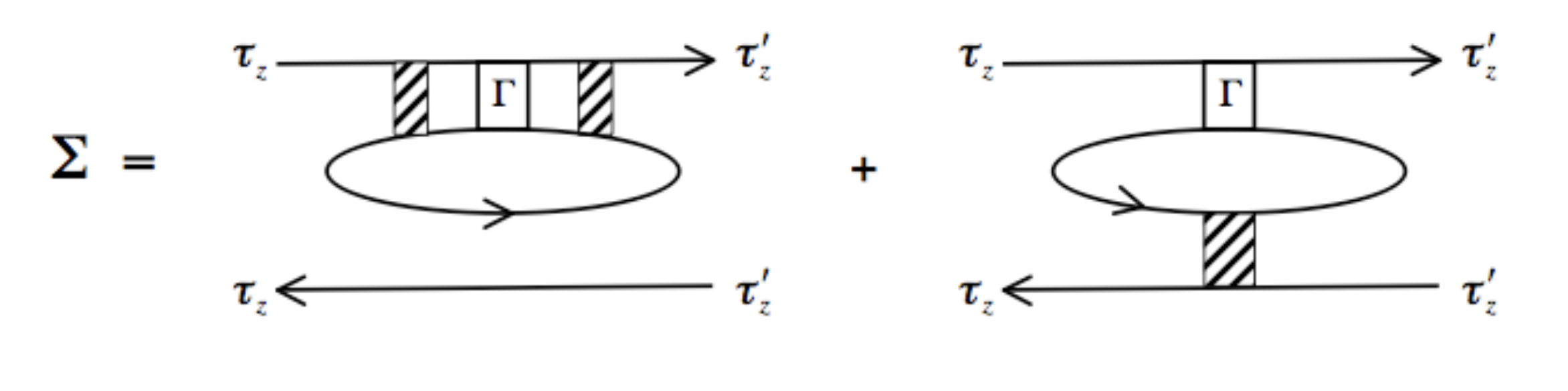}
\caption{The skeleton diagrams for  $\Sigma$ are listed. $\Gamma$ is the interaction matrix  and the hashed blocks are diffusion propagators. The $\Sigma_+$ matrix is obtained by adding the contributions from $\uptau_z'=\pm \uptau_z$. The results are presented in Eq.\;(\ref{eqn:sigma_skeleton}).
}  \label{fig:sigma_skeleton}
\end{figure}

The \textit{e-e} interaction corrections to the diffusion propagators are expressed in terms of the ``self-energy" matrix $\Sigma$. The relevant diagrams are shown in Fig.\;\ref{fig:sigma_skeleton}. Expanding $\Sigma(q,\omega)$ to order $q^2$ and $\omega$ one obtains, for example, for the gapless $\mathcal{D}_+$ propagator, the renormalized propagator  
$\mathcal{D}^{-1}_+(q,\omega)=Dq^2+z\omega$, where $D$ is the renormalized diffusion constant  and $z$ is the frequency renormalization parameter that determines the change in the relative scaling of the frequency with respect to the length scale\;\cite{sasha83,sasha84} ($z=1$ for non-interacting electrons).  The corresponding corrections to $D$ and $z$ obtained by evaluating the diagrams in Fig.\;\ref{fig:sigma_skeleton} are given in Eq.\;(\ref{eqn:sigma_skeleton}) in Appendix\;\ref{app:corrections}.


\begin{figure}[b]
\includegraphics[width=0.5\linewidth]{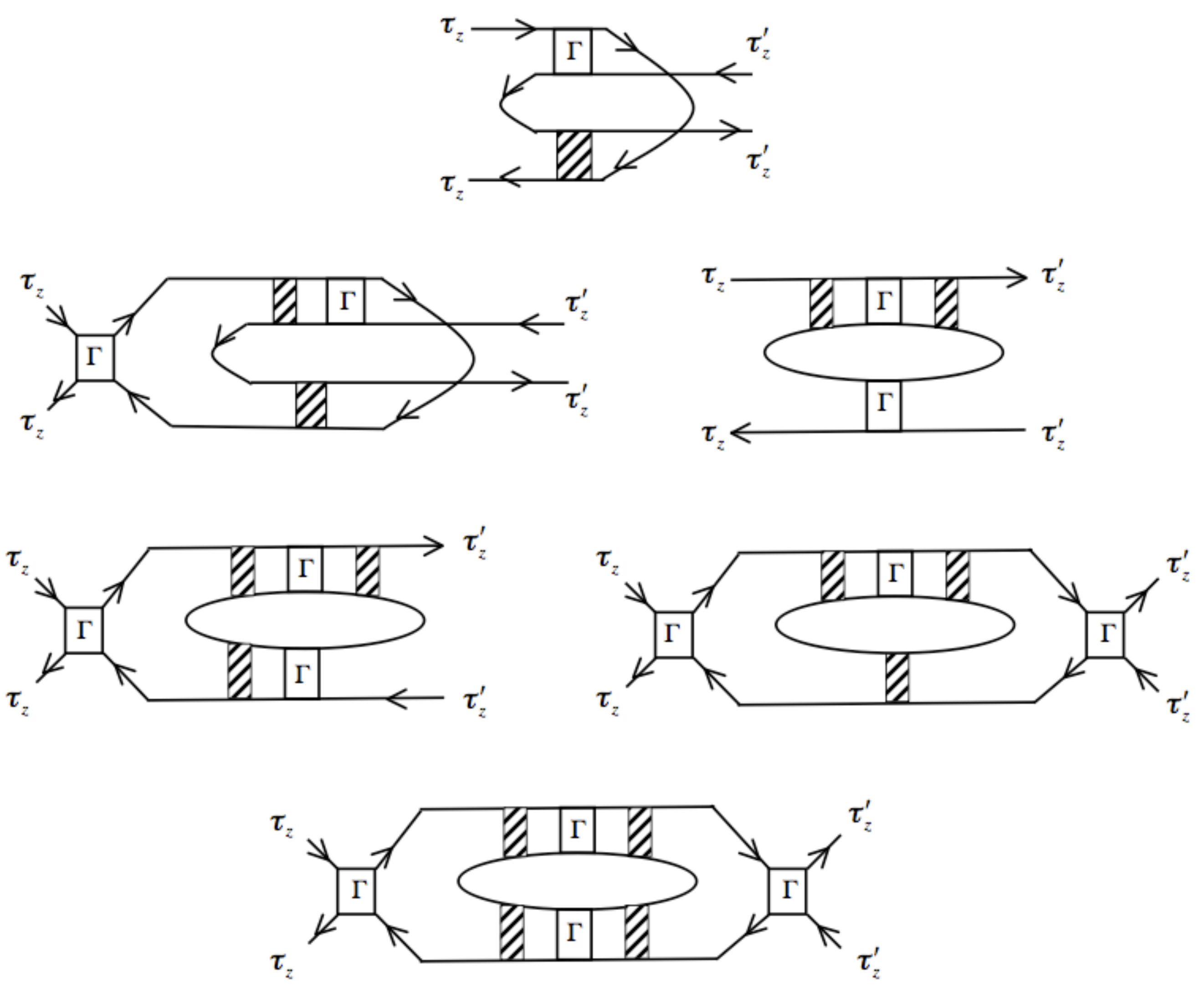}
\caption{Skeleton diagrams for $\delta\Gamma_{i,\alpha}$. By appropriately choosing the $\Gamma$ vertices  for given values of $\uptau_z'=\pm\uptau_z$ all the corrections $\delta\Gamma_{i,\alpha}$, where $i=1,2$ and $\alpha=\pm$, are calculated. Each combination of the valley indices comes together with the appropriate diffuson matrix elements. Note that going down the rows increases the   number of $\Gamma$ vertices.  }  \label{fig:gamma_skeleton}
\end{figure}

 The  skeleton-diagrams representing  the diffusion corrections to the \textit{e-e} scattering amplitudes are shown in Fig.\;\ref{fig:gamma_skeleton}. (For a detailed discussion of these corrections, see Refs.\;\cite{yellowbook,pedagogical}.)
The calculations are generalized here to include valleys. By appropriately choosing the $\Gamma$ vertices  for given values of $\uptau_z'=\pm\uptau_z$ in Fig.\;\ref{fig:gamma_skeleton}  all the corrections, $\delta\Gamma_{i,\alpha}$, to the scattering amplitudes $\Gamma_{i,\alpha}$, where $i=1,2$ and $\alpha=\pm$, can be calculated. %
For example, to calculate $\delta\Gamma_{i+}$, since $\Gamma_{i+}=\Gamma_{i,\parallel,u}+\Gamma_{i,\parallel,v}$, the contributions from $\uptau_z'=\pm\uptau_z$ are added, while they are subtracted when calculating $\delta\Gamma_{i-}$.
The results  are given in Eq.\;(\ref{eqn:gamma_skeleton}) in Appendix\;\ref{app:corrections}. (The corrections to the amplitude $\delta\Gamma_\perp$ are not given as they are equal to $\delta\Gamma_{2+}$ for $T\agt T_v$ and irrelevant for $T\alt T_v$ due to the gap.) 

The corrections $\delta D$, $\delta z$ and $\delta \Gamma_{i,\alpha}$ in Eqs.\;(\ref{eqn:sigma_skeleton}) and (\ref{eqn:gamma_skeleton}) include all modes, both gapped and gapless. 
Clearly, only modes that are effectively gapless lead to logarithmically divergent corrections. Since the frequency integrations range from $T\leq \omega \leq 1/\tau$ (the upper cut-off follows from taking the diffusion limit), for $T \alt T_*$, both $\mathcal{D}_-$ and $\mathcal{D}_\perp$ are gapped, while only the $\mathcal{D}_-$ modes are effectively gapless when $T_*\alt T\alt T_v$. (The $\mathcal{D}_+$ mode is always gapless.) Of course, when $T\agt T_v$ and $T_*$, all modes are gapless. As as result, the corrections are clearly sensitive to the temperature range considered. 

\subsection{High temperature range: $T\agt T_v$ and $T_*$}\label{sec:corrections_highT}

For $T\agt T_v$ and $T_*$, all the modes $\mathcal{D}_\alpha$ ($\alpha=\pm,\perp$) appearing in Eqs.\;(\ref{eqn:sigma_skeleton}) and (\ref{eqn:gamma_skeleton}) are effectively gapless, i.e., they take the form  $\mathcal{D}(q,\omega)=1/(Dq^2+z\omega)$. 
As noted below, not all amplitudes $\Gamma_{i,\alpha}$ are relevant at these temperatures. For instance, since intervalley scattering is irrelevant for $T\agt T_*$, the amplitudes  $\Gamma_{1\perp}$ and $\Gamma_{2,\parallel,v}$, whose initial values are vanishingly small, can be set to zero. As a result (see Fig.\;\ref{fig:gamma_skeleton}),  $\Gamma_{1\perp}\approx 0$ and $\Gamma_{2+}\approx\Gamma_{2-}$.  Further more, since valley splitting can be ignored for $T\agt T_v$, the amplitudes $\Gamma_{1,\parallel,u}$ and $\Gamma_{2\perp}$ are indistinguishable from the amplitudes  $\Gamma_{1,\parallel,v}$ and $\Gamma_{2+}$, respectively, implying that the initial value of $\Gamma_{1-}=0$ and  $\Gamma_{2\perp}=\Gamma_{2+}$.

It can be seen from Eq.\;(\ref{eqn:gamma_skeleton}) that choosing the above initial conditions, namely, $\Gamma_{1\perp}=\Gamma_{1-}=0$, and setting all the $\Gamma_{2\alpha}$ amplitudes to be equal,  and all the $\mathcal{D}_\alpha$ propagators to be gapless, gives $\delta\Gamma_{1-}=0$ and $\delta\Gamma_{2-}=\delta\Gamma_{2+}$, which are consistent with the choice of the initial conditions. Hence,  Eqs.\;(\ref{eqn:sigma_skeleton}) and (\ref{eqn:gamma_skeleton}) reduce to the form (with the substitution  $\Gamma_{2\alpha}\equiv\Gamma_2$ and $\mathcal{D}_\alpha\equiv\mathcal{D}$):

\begin{subequations}\label{eqn:corrections_highT}
\begin{eqnarray}
\frac{\delta D}{D}&=&-\frac{4}{\nu}\iint \frac{d\omega}{2\pi} \left(\Gamma_{1+}-4\Gamma_2\right)\mathcal{D}^3(q,\omega)Dq^2\\
\delta z&=&-\frac{1}{\pi\nu}\int\frac{d^2q}{(2\pi)^2} \left(\Gamma_{1+}-4\Gamma_2\right)\mathcal{D}(q,0)\\
\delta\Gamma_{1+}&=&\frac{1}{\pi\nu}\int\frac{d^2q}{(2\pi)^2}\Gamma_2\mathcal{D}(q,0)+4\Psi(\Gamma_2)\\
\delta\Gamma_2&=&\frac{1}{\pi\nu}\int\frac{d^2q}{(2\pi)^2}\Gamma_{1+}\mathcal{D}(q,0)+16\Psi(\Gamma_2)
\end{eqnarray}
\end{subequations}
where $\iint=\int dq^2/(2\pi)^2\int d\omega/(2\pi)$ and $\Psi(\Gamma_2)$ equals
\begin{eqnarray}\label{eqn:Psi}
\Psi(\Gamma_2)&=&
+\frac{1}{\nu}\int\frac{d^2q}{(2\pi)^2}\int \frac{d\omega}{2\pi}\Gamma_2\left[\Gamma_2\mathcal{D}^2\right]-\frac{1}{2}\left[\Gamma_2^2\mathcal{D}^2\right]\nonumber\\
&&-\frac{1}{\nu}\int\frac{d^2q}{(2\pi)^2}\int \frac{d\omega}{2\pi}\omega\Gamma_2
\left[\Gamma_2^2\mathcal{D}^3\right]-\omega\Gamma_2^2\left[\Gamma_2\mathcal{D}^3\right]\nonumber\\
&&-\frac{2}{\nu}\int\frac{d^2q}{(2\pi)^2}\int \frac{d\omega}{2\pi}\omega^2\Gamma_2^2\left[\Gamma_2^2\mathcal{D}^4\right]
\end{eqnarray}

The above equations  were first obtained in Ref.\;\cite{punnoose02}, they correspond to  the case when the two valleys are distinct and degenerate.

\subsection{Low temperature range: $T\alt T_*$ }\label{sec:corrections_lowT}

When $T\alt T_*$, both $\mathcal{D}_-$ and $\mathcal{D}_\perp$ are gapped and therefore irrelevant. Hence, only the $\mathcal{D}_+$ mode survives. Dropping the contributions of the gapped modes in  Eqs.\;(\ref{eqn:sigma_skeleton}) and (\ref{eqn:gamma_skeleton}) lead to a self contained set of equations involving only the amplitudes $\delta\Gamma_{1+}$ and $\delta\Gamma_{2+}$. The equations, after dropping the $+$ sign in $\mathcal{D}_+$ and $\Gamma_{2+}$, reduce to
\begin{subequations}\label{eqn:corrections_lowT}
\begin{eqnarray}
\frac{\delta D}{D}&=&-\frac{4}{\nu}\iint \left(\Gamma_{1+}-\Gamma_2\right)\mathcal{D}^3(q,\omega)Dq^2\\
\delta z&=&-\frac{1}{\pi\nu}\int\frac{d^2q}{(2\pi)^2} \left(\Gamma_{1+}-\Gamma_2\right)\mathcal{D}(q,0)\\
\delta\Gamma_{1+}&=&\frac{1}{4\pi\nu}\int\frac{d^2q}{(2\pi)^2}\Gamma_2\mathcal{D}(q,0)+\Psi(\Gamma_2)\\
\delta\Gamma_2&=&\frac{1}{\pi\nu}\int\frac{d^2q}{(2\pi)^2}\Gamma_{1+}\mathcal{D}(q,0)+{4}\Psi(\Gamma_2)
\end{eqnarray}
\end{subequations}

These equations correspond to the case when the two valleys appear as a single valley due to intervalley scattering.  (Note that valley splitting is  irrelevant in this case as the $\mathcal{D}_\perp$ propagator is always gapped when $T\alt T_*$, irrespective of $T_v$.)  

\subsection{Intermediate temperature range: $T_*\alt T\alt T_v$}\label{sec:corrections_medT}

This limit when the valley splitting is large, so that the intervalley scattering rate $T_*\ll T_v$, is interesting.  For temperatures in the intermediate range $T_*\alt T\alt T_v$, only the $\mathcal{D}_\perp$ mode is gapped, while both $\mathcal{D}_\pm$ are gapless.   Although the initial value of $\Gamma_{1-}\approx 0$ when $T\agt T_v$ (see discussion in Sec.\;\ref{sec:corrections_highT}) it can be seen from Eq.\;(\ref{eqn:gamma_skeleton}) that $\delta\Gamma_{1-}\neq 0$ when $T\alt T_v$ and is therefore generated at intermediate temperatures.   This introduces a third relevant scaling parameter distinct from the high and low temperature regimes. (Since $T\agt T_*$,  $\Gamma_{2+}=\Gamma_{2-}$, but because $T\alt T_v$ the  $\Gamma_{2\perp}$ amplitude is irrelevant.)

Dropping the $\mathcal{D}_\perp$ terms in  Eqs.\;(\ref{eqn:sigma_skeleton}) and (\ref{eqn:gamma_skeleton}) and setting  $\Gamma_{2+}=\Gamma_{2-}\equiv \Gamma_2$ and $\mathcal{D}_\pm=\mathcal{D}$  gives
\begin{subequations}\label{eqn:corrections_medT}
\begin{eqnarray}
\frac{\delta D}{D}\!\!&=&\!\!-\frac{4}{\nu}\iint \left(\Gamma_{1-}+\Gamma_{1+}-2\Gamma_2\right)\mathcal{D}^3(q,\omega)Dq^2\hspace{0.5cm}\\
\delta z&=&-\frac{1}{\pi\nu}\int\frac{d^2q}{(2\pi)^2} \left(\Gamma_{1-}+\Gamma_{1+}-2\Gamma_2\right)\mathcal{D}(q,0)\\
\delta\Gamma_{1+}\!\!&=&\!\!\delta\Gamma_{1-}=\frac{1}{2\pi\nu}\int\frac{d^2q}{(2\pi)^2}\Gamma_2\mathcal{D}(q,0)+{2}\Psi(\Gamma_2)\\
\delta\Gamma_2\!\!&=&\!\!\frac{1}{\pi\nu}\int\frac{d^2q}{(2\pi)^2}\left(\Gamma_{1+}+\Gamma_{1-}\right)\mathcal{D}(q,0)+{8}\Psi(\Gamma_2)\hspace{0.85cm}\end{eqnarray}
\end{subequations}
 Note that although both $\delta \Gamma_{1+}$ and $\delta\Gamma_{1-}$ are equal, their initial values are different.

The relevance of the $\Gamma_{1-}$ amplitude in the temperature range  $ T_*\alt T\alt T_v$ is specific to problems with split-bands, and was  first discussed  in Ref.\;\cite{burmistrov_spinvalley} for the case of spin-splitting in a multi-valley system.

\section{Renormalization group equations}

\begin{figure}[htb]
\includegraphics[width=0.5\linewidth]{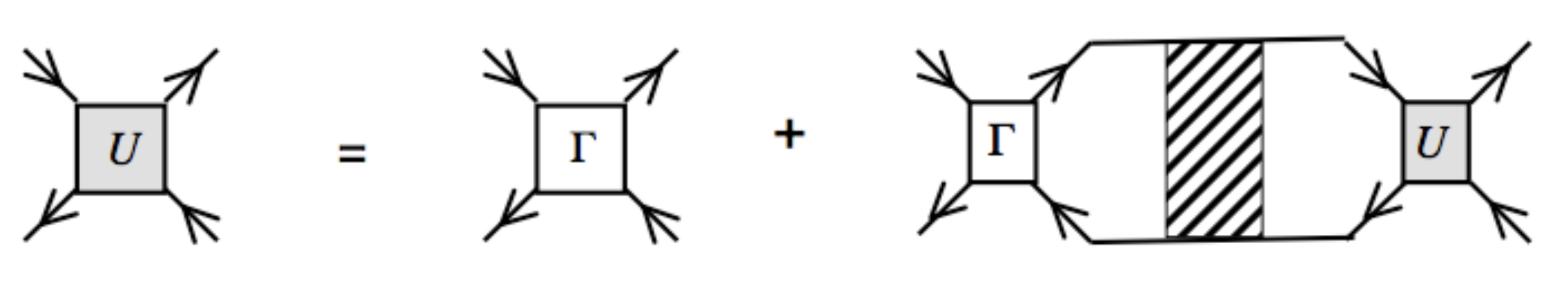}
\caption{Extending the static amplitudes $\Gamma$ by the corresponding dynamic amplitudes $U$ through ladder summations. }  \label{fig:ladder}
\end{figure}

In  Sec.\;\ref{sec:corrections}, the leading logarithmic corrections in all the different temperature ranges have been listed. It is now possible to set up the scaling equations. To this end, first note that all the corrections involve only one momentum integration, and since every momentum integration generates a factor of $1/D$, which by Einstein's relation is proportional to the resistance $\rho$, the corrections are limited to the first order in resistance (disorder). The limitation on the number of momentum integrations also constraints the number of \textit{e-e} vertices in the skeleton diagrams shown in Figs.\;\ref{fig:sigma_skeleton} and \ref{fig:gamma_skeleton}.  These corrections can now be extended to all orders in $\Gamma$ (but still first order in $\rho$) by performing  ladder summations as shown in Fig.\;\ref{fig:ladder}.  It amounts to replacing the static  amplitudes $\Gamma$ by the dynamical amplitudes $U(q,\omega$) as discussed below.


Since the ladder summations  do not introduce additional momentum integrations, the resummation allows the  corrections to be evaluated to infinite order in the interaction amplitude leaving $\rho$ as the only expansion parameter in the theory\;\cite{sasha83}.

For the amplitudes $\Gamma_{2\alpha}$, the ladder sums are most easily done in the basis $\alpha=\pm$ and $\perp$,  as it can be checked by inspection that  the indices are conserved in the ladder. Using $\Gamma_{2\alpha}$ and $\mathcal{D}_\alpha$ in Fig.\;\ref{fig:ladder}, one obtains the corresponding dynamical amplitude $U_{2\alpha}(q,\omega)$, where
\begin{equation}\label{eqn:U2}
U_{2\alpha}(q,\omega)=\Gamma_{2\alpha} \frac{\mathcal{D}_{2\alpha}(q,\omega)}{\mathcal{D}_\alpha(q,\omega)}
\end{equation}
The propagators $\mathcal{D}_\alpha$ are defined in Eq.\;(\ref{eqn:diffusons}) and 
\begin{equation}
\mathcal{D}_{2\alpha}(q,\omega)=\frac{1}{Dq^2+(z+\Gamma_{2\alpha})\omega+\Delta_\alpha}
\end{equation}

It should be noted in, for example, Fig.\;\ref{fig:gamma_skeleton}, that only those interaction vertices involving frequency integrations can be extended to include dynamical effects. For convenience, the corresponding $\Gamma_2$ vertices are enclosed in square brackets in the function $\Psi$ in Eq.\;(\ref{eqn:Psi}). Substituting for $\Gamma_2$ in Eq.\;(\ref{eqn:Psi}) with $U_2$ from Eq.\;(\ref{eqn:U2}) (the $\alpha$ index is dropped since only gapless modes have been retained in (\ref{eqn:Psi})) and performing the $q$ and $\omega$ integrals leads to the very simple expression\;\cite{sasha83,pedagogical}:
\begin{equation}\label{eqn:psi_integral}
\Psi(\Gamma_2)=\left(\frac{\Gamma_2^2}{z}\right)\times \frac{\rho}{2}\log\left(\frac{1}{T\tau}\right)
\end{equation} 
The dimensional resistance $\rho=1/4(2\pi^2\nu D)$ corresponds to $(e^2/\pi h)R_\square$, where $R_\square$ is the sheet resistance. The factor $4$ arises due to the spin and valley degrees of freedom and $\nu$ is the density of states per spin and valley. Also note that up to logarithmic accuracy the upper cut-off can be replaced with $1/\tau$.
Since, the remaining integrals in Eqs.\;(\ref{eqn:corrections_highT}) to (\ref{eqn:corrections_medT}) are of the form $\int d^2q \mathcal{D}(q,0)$, they can be evaluated directly as
\begin{equation}\label{eqn:rhointegral}
\frac{1}{\pi\nu}\int \frac{d^2q}{(2\pi)^2}\mathcal{D}(q,0)=2\rho\log\left(\frac{1}{T\tau}\right)
\end{equation}

It remains to evaluate the integrals for $\delta D$ and $\delta z$. The $\delta z$ integrals do not involve frequency integrations and can therefore be evaluated using Eq.\;(\ref{eqn:rhointegral}). The $\delta D$ corrections, however, contain frequency integrals, and therefore the $\Gamma_{1\pm}$ amplitudes, in addition to  $\Gamma_2$,  are also to be extended to all orders via the ladder sum. 

This is most easily done in the  spin-singlet basis 
\begin{equation}\label{eqn:singlets}
\Gamma_{s\pm}=\Gamma_{1\pm}-\frac{1}{4}\Gamma_{2\pm}
\end{equation} 
This is so, because the spin and valley of the electron-hole pairs in the singlet and triplet basis are individually conserved in the ladder sum. (Note that the `$+$' amplitude is written in the (spin-singlet)$\otimes$(valley-singlet) basis, while the `$-$' amplitude is in the (spin-singlet)$\otimes$(valley-triplet) basis; the valley-triplet corresponds to $|S=1,S_z=0\rangle$.)
The  corresponding dynamical amplitudes $U_{s\pm}(q,\omega)$ on performing the ladder sum gives 
\begin{equation}\label{eqn:dynamical_singlet}
U_{s\pm}(q,\omega)=\Gamma_{s\pm}\frac{\mathcal{D}_{s\pm}(q,\omega)}{\mathcal{D}_\pm(q,\omega)}
\end{equation}
where
\begin{equation}
\mathcal{D}_{s\pm}(q,\omega)=\frac{1}{Dq^2+(z-4\Gamma_{s\pm})\omega+\Delta_\pm}
\end{equation}
(Note that $\Delta_+$ is introduced for notational uniformity, in fact $\Delta_+=0$.)

\begin{figure}[htb]
\includegraphics[width=0.25\linewidth]{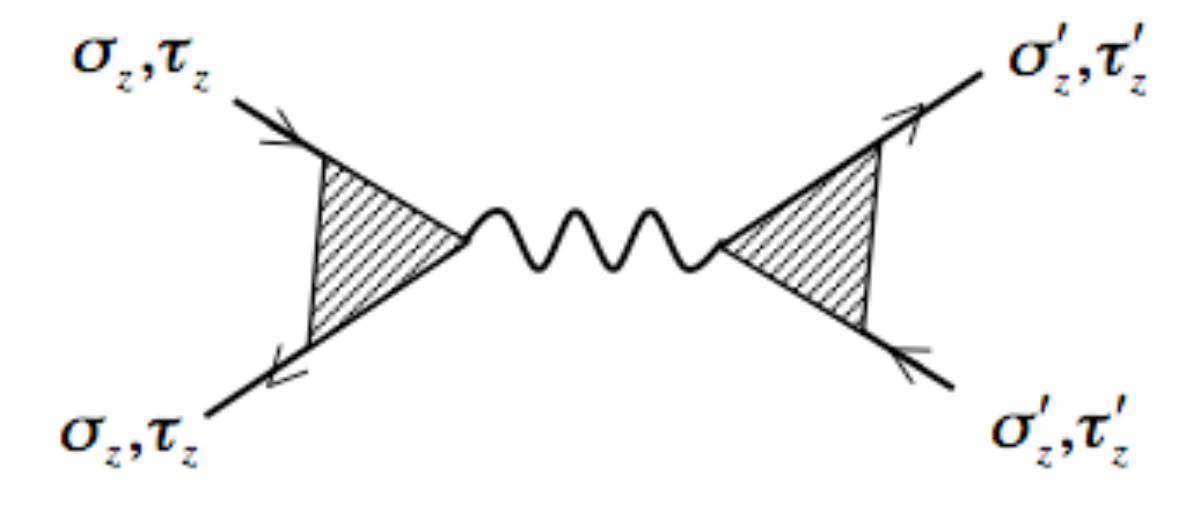}
\caption{The statically screened long range part of the Coulomb interaction, which can be separated by cutting just one Coulomb line is shown. The amplitude $\Gamma_{0+}$ is obtained by adding the amplitudes with $\uptau'_z=\pm$.   The shaded triangles represent the static vertex corrections $V$. }  \label{fig:gamma0}
\end{figure}

Special attention is to be paid to the ladder sums involving $\Gamma_{s+}$  when  Coulomb interactions are present. The $\Gamma_{1+}$ amplitudes in this case includes amplitudes of the kind shown in Fig.\;\ref{fig:gamma0}, which can be separated by cutting the statically screened long-ranged Coulomb line once. They are denoted here as $\Gamma_{0+}$.  This distinction is important because the polarization operator, $\Pi(q,\omega)$, which is irreducible to cutting a Coulomb line does not include $\Gamma_{0+}$. (The corresponding $\Gamma_{0-}$ and $\Gamma_{0\perp}$ amplitudes are zero. The former is identically zero, while the latter involving intervalley scattering is vanishingly small.)

Analyzing the polarization operator, $\Pi(q,\omega)$, provides key insights into the relationship between the various amplitudes (and $z$)\;\cite{yellowbook,pedagogical}. The form of $\Pi(q,\omega)$ is analyzed here in the presence of valleys. 
In the limit of $q,\omega\rightarrow 0$, it can be shown that $\Pi(q,\omega)$  takes the form:
\begin{equation}
\Pi(q,\omega)=-\frac{\partial n}{\partial \mu}+ \frac{4V^2\omega}{Dq^2+\left(z-4\Gamma_{s+}\right)\omega}\label{eqn:pi0}
\end{equation}
It is important to note that  only the $\Gamma_{s+}$ amplitude, corresponding to the singlet mode, appears in the expression for $\Pi(q,\omega)$.  The factor $\partial n/\partial \mu$ is the thermodynamic density of states and the parameter $V$ is the static vertex corrections represented as shaded triangles in Fig.\;\ref{fig:gamma0}. 

The two terms in Eq.\;(\ref{eqn:pi0}) correspond to the static and the dynamical contributions, respectively. By construction, the static limit $\Pi(q\rightarrow 0,\omega=0)=-\partial n/\partial \mu$ is satisfied.  In the opposite limit, local conservation law requires that $\Pi(q=0,\omega\rightarrow 0)=0$.  From Eq.\;(\ref{eqn:pi0}) it can be seen that for the latter condition to be satisfied the following relation must hold:
\begin{equation}
\frac{\partial n}{\partial \mu}=\frac{4V^2}{z-4\Gamma_{s+}}\label{eqn:dos}
\end{equation}
in which case, $\Pi(q,\omega)$ takes the form:
\begin{equation}
\Pi(q,\omega)=-\frac{\partial n}{\partial \mu}\times\frac{Dq^2}{Dq^2+(z-4\Gamma_{s+})\omega}\label{eqn:pi}
\end{equation}

When Eq.\;(\ref{eqn:dos}) is combined with the definition of $\Gamma_{0+}$ as the static limit of the Coulomb interaction, i.e., $\Gamma_{0+}=V^2{\partial \mu}/{\partial n}$, the following expression for $\Gamma_{0+}$ is obtained: 
%
$\Gamma_{0+}=\frac{1}{4}(z-4{\Gamma}_s).$ Hence, conservation laws provide the  
%
very important relation
%
\begin{equation}
z=4\left(\Gamma_{0+}+\Gamma_{1+}\right)-\Gamma_{2+}\equiv 4\Gamma_{s+}^{LR}\label{eqn:z}
\end{equation}
where $\Gamma_{s+}^{LR}=\Gamma_{1+}^{LR}-\Gamma_{2+}/4$,  denotes the singlet amplitude in the presence of long-ranged Coulomb interactions. Since, only the Coulomb case is considered in the following, all the $\Gamma_{1+}$ amplitudes appearing in Eqs.\;(\ref{eqn:corrections_highT}) to (\ref{eqn:corrections_medT}) are to be replaced by their long-ranged counterparts
\begin{equation}\label{eqn:gamma1_LR}
\Gamma_{1+}\longrightarrow\Gamma^{LR}_{1+}=\Gamma_{0+}+\Gamma_{1+}
\end{equation}

Direct inspection of Eqs.\;(\ref{eqn:corrections_highT}) to (\ref{eqn:corrections_medT}) shows that the singlet combination in Eq.\;(\ref{eqn:z}) is satisfied everywhere, i.e.,  $\delta (z - 4\Gamma_{s}^{LR})=0$, provided $\delta \Gamma_{0+}=0$. (This is a well established result, with great importance for the general structure of the theory\;\cite{sasha83,yellowbook}.) 
In particular, the corresponding dynamical amplitude $U^{LR}_{s+}(q,\omega)$ reads 
\begin{equation} \label{eqn:Us}
U^{LR}_{s+}(q,\omega)=\frac{z}{4Dq^2}\frac{1}{\mathcal{D}(q,\omega)}
\end{equation} 
Note that unlike the $U_{s+}$ amplitude (see \ref{eqn:dynamical_singlet})  $U^{LR}_{s+}$ is  a universal amplitude independent of $\Gamma^{LR}_{s+}$. This is a direct consequence of the singlet relation (\ref{eqn:z})\;\cite{aabook,sasha83}.

The scaling equations discussed below are obtained from Eqs.\;(\ref{eqn:corrections_highT}) to (\ref{eqn:corrections_medT}) after  (i)\;rearranging all the $\Gamma_{1\pm}$ amplitudes to give $\Gamma_{s\pm}$, and then replacing $\Gamma_{s+}$ with $\Gamma_{s+}^{LR}$, (ii)\;replacing the static amplitudes  where applicable by the corresponding dynamical amplitudes and  (iii)\;substituting $\rho=1/4(2\pi^2\nu D)$. 

It is convenient to express the equation for $\rho$ in terms of the scaling variables,  $\gamma_2=\Gamma_2/z$ and $\gamma_v=-4\Gamma_{s-}/z$. In terms of these variables, the equations for  $\rho$, $\gamma_2$ and $\gamma_v$ form a closed set of  equations  independent of $z$. The final RG equations, along with the equations for $z$, are given below. The scale $\xi=\log(1/T\tau)$ is used in these equations. To logarithmic accuracy $1/\tau$ can be used as the upper cut-off. The range of applicability of $\xi$ is defined in each case separately.

\begin{enumerate}
\item\textit{High temperature limit}: $T\agt T_v$ and $T_*$
\begin{subequations}\label{eqn:RG_highT}
\begin{eqnarray}
\frac{d\rho}{d\xi}&=& \rho^2\left(1-15\Phi(\gamma_2)\right)\\
\frac{d\gamma_2}{d\xi}&=&\frac{\rho}{2}(1+\gamma_2)^2\\
\frac{d\ln z}{d\xi}&=&-\frac{\rho}{2}\left(1-15\gamma_2\right)
\end{eqnarray}
\end{subequations}
\item\textit{Low temperature limit}: $T\alt T_*$
\begin{subequations}\label{eqn:RG_lowT}
\begin{eqnarray}
\frac{d\rho}{d\xi}&=& \rho^2\left(1-3\Phi(\gamma_2)\right)\\
\frac{d\gamma_2}{d\xi}&=&\frac{\rho}{2}(1+\gamma_2)^2\\
\frac{d\ln z}{d\xi}&=&-\frac{\rho}{2}\left(1-3\gamma_2\right)
\end{eqnarray}
\end{subequations}
\item\textit{Intermediate temperature limit}: $T_*\alt T\alt T_v$
\begin{subequations}\label{eqn:RG_medT}
\begin{eqnarray}
\frac{d\rho}{d\xi}&=& \rho^2\left(1-\Phi(\gamma_v)-6\Phi(\gamma_2)\right)\\
\frac{d\gamma_2}{d\xi}&=&\frac{\rho}{2}\left[(1+\gamma_2)^2+(1+\gamma_2)(\gamma_2-\gamma_v)\right]\\
\frac{d\gamma_v}{d\xi}&=&\frac{\rho}{2}(1+\gamma_v)(1-\gamma_v-6\gamma_2)\\
\frac{d\ln z}{d\xi}&=&-\frac{\rho}{2}\left(1-\gamma_v-6\gamma_2\right)
\end{eqnarray}
\end{subequations}

\end{enumerate}
The variable $\Phi(\gamma)$ is defined as
\begin{equation}
\Phi(\gamma)=\left(1+\frac{1}{\gamma}\right)\log(1+\gamma)-1
\end{equation} 

The factors $15$ and $3$ appearing in Eqs.\;(\ref{eqn:RG_highT}) and (\ref{eqn:RG_lowT}), respectively, correspond to the number of effective triplet modes. In the case of two distinct, degenerate valleys, the 16 spin-valley modes break up into $1$ singlet and $15$ ``triplet" modes, while in the limit of strong intervalley scattering the two valleys are effectively combined into a single valley leading to $3$ spin-triplet modes. 

When the valleys are split, as in Eq.\;(\ref{eqn:RG_medT}), the amplitude $\gamma_v$ plays a significant role as the temperature is reduced well below $T_v$. Given that $\gamma_v=(\Gamma_2-4\Gamma_{1-})/z$ and that $\Gamma_{1-}\approx 0$ for  $T \agt T_v$, it follows that  $\gamma_v\approx \gamma_2$ as $T$ approaches $T_v$ from above. When $T\ll T_v$, the two amplitudes $\gamma_2$ and $\gamma_v$ diverge from each other significantly. For $T\alt T_v$, however, it is reasonable to assume that $\gamma_v\approx \gamma_2$. This is relevant if the lower cut-off $T_*$ is not much smaller than $T_v$. In this case, the equation for $\rho$ and $\gamma_2$ pertaining to the different temperature ranges can be combined  to give
$d\rho/d\xi=\rho^2(1-(4K-1)\Phi(\gamma_2))$, and $d\gamma_2/d\xi=\rho(1+\gamma_2)^2/2$. Here, $K=n_v^2=4$ when the valleys are degenerate and distinct (high temperature), $K=n_v^2=1$ when intervalley scattering is strong (low temperature) and $K=n_v=2$ when the valleys are distinct but split so that each valley contributes independently (intermediate temperature). For direct comparison with experiments,  these simplified equations should suffice for most samples. 

The situation changes, however, once $T \ll T_v$, but still greater than $T_*$.  We see that $\gamma_v$ and $\gamma_2$ evolve differently until $\gamma_v$ reaches the fixed point value of $\gamma_v^*=-1$, at which point $\Phi(-1)=-1$. (This fixed point is relevant only when $T_*\approx 0$.) The system at this point reduces to a single valley system with resistance $2\rho$. The above properties are generic to systems with split bands (spin and valley) as has been discussed in detail in Ref.\;\cite{burmistrov_spinvalley}.

To summarize, RG equations have been obtained in the case when both valley splitting and intervalley scattering are present.  The results  can be directly used to compare with experiments  in a two-valley system after adding the weak-localization contributions, which are not included here. The case when the two bands are split but otherwise distinct is quantitatively different due to the existence of a third relevant scaling parameter. The asymptotic metallic behavior is, however, not affected.   

\section{Acknowledgments}

The author would like to thank A.\ M.\ Finkel'stein for numerous discussions on this topic. This work was supported by  DOE grant DOE-FG02-84-ER45153 and US-Israel Binational Science Foundation grant 2006375. 

\appendix


\section{Diffusion propagators}\label{app:diffuson}

\begin{figure}[htb]
\includegraphics[width=0.5\linewidth]{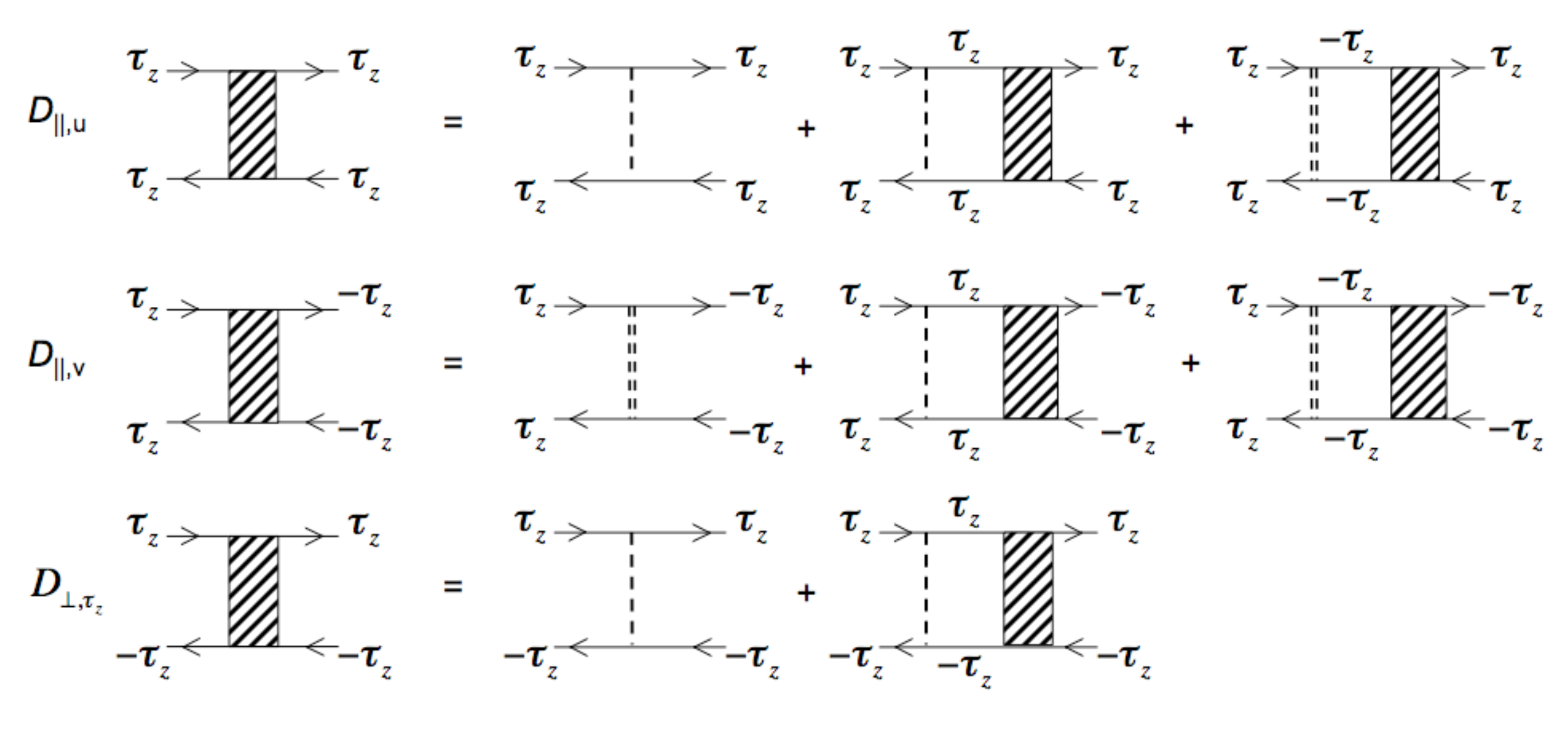}
\caption{Ladder diagrams for the various diffuson blocks are shown above. As in Fig.\;\ref{fig:tau}, the intra and intervalley impurity scatterings are denoted by single and double dashed lines, respectively. The particle and hole Greens functions for finite $\Delta_v$ correspond to $\mathcal{G}_{\uptau_z}(k,\epsilon)=[i\epsilon-(\xi_k-\uptau_z \Delta_v/2) + i/2\tau\; \textrm{sgn}\,\epsilon]^{-1}$}  \label{fig:app_diffuson_blocks}
\end{figure}

The ladder diagrams for each of the diffuson blocks,  $\mathcal{D}_{\parallel,u}, \mathcal{D}_{\parallel,v}$ and $\mathcal{D}_{\perp,\uptau_z}$, are detailed in Fig.\;\ref{fig:app_diffuson_blocks}. The corresponding equations are given in Eq.\;(\ref{eqn:app_d_ladder}).
Note that the $\mathcal{D}_\parallel$ diffusons are coupled in the presence of intervalley scattering. For convenience, the scattering rates  in Eq.\;(\ref{eqn:uv}) are defined as $\Delta_\parallel=1/2\pi\nu\tau_\parallel$ and $\Delta_\perp=1/2\pi\nu\tau_\perp$. 
\begin{subequations}
\begin{eqnarray}
\mathcal{D}_{\parallel,u}&=&\Delta_\parallel+\Delta_\parallel X_\parallel\mathcal{D}_{\parallel,u}+\Delta_\perp X_\parallel\mathcal{D}_{\parallel,v}\label{eqn:app_du}\\
\mathcal{D}_{\parallel,v}&=&\Delta_\perp+\Delta_\parallel X_\parallel\mathcal{D}_{\parallel,v}+\Delta_\perp X_\parallel\mathcal{D}_{\parallel,u}\label{eqn:app_dv}\\
\mathcal{D}_{\perp,\uptau_z}&=&\Delta_\parallel+\Delta_\parallel X_{\perp,\uptau_z}\mathcal{D}_{\perp,\uptau_z}\label{eqn:app_dperp}
\end{eqnarray}\label{eqn:app_d_ladder}
\end{subequations}
where
\begin{subequations}
\begin{eqnarray}
X_\parallel(q,\omega)&=&\sum_\mathbf{k}\mathcal{G}_{\uptau_z}(\mathbf{k}+\mathbf{q},\epsilon+\omega)\mathcal{G}_{\uptau_z}(\mathbf{k},\epsilon)\nonumber\\
&\approx&2\pi\nu\tau\left[1-\tau\omega-\tau D_0q^2+\cdots\right]\\
X_{\perp,\uptau_z}(q,\omega)&=&\sum_\mathbf{k}\mathcal{G}_{\uptau_z}(\mathbf{k}+\mathbf{q},\epsilon+\omega)\mathcal{G}_{-\uptau_z}(\mathbf{k},\epsilon)\nonumber\\
&\approx&2\pi\nu\tau\left[1-\tau\omega-\tau D_0q^2+i\uptau_z (\Delta_v\tau)\right]\label{eqn:app_Xperp}\hspace{0.75cm}
\end{eqnarray}\label{eqn:app_X}
\end{subequations}
Here $D$ is the diffusion constant.  In the diffusion approximation, i.e., for $\epsilon(\epsilon+\omega)<0$, it is sufficient to evaluate $X$ in the long wavelength and small frequency limit. Only the weak splitting $\Delta_v\tau \lesssim 1$ limit is considered here. 

Eqs.\;(\ref{eqn:app_du}) and (\ref{eqn:app_dv}) are easily decoupled by defining 
\begin{equation}
\mathcal{D}_\pm=\mathcal{D}_{\parallel,u}\pm\mathcal{D}_{\parallel,v}=\frac{1}{2\pi\nu\tau^2}\times\frac{1}{D_0q^2+\omega+\Delta_\pm}\label{eqn:app_dpm}
\end{equation}
where $\Delta_+=0$ and $\Delta_-=2\tau_\parallel/\tau(\tau_\perp-\tau_\parallel)$. Note that $\mathcal{D}_+$ is gapless. Substituting (\ref{eqn:app_Xperp}) in (\ref{eqn:app_dperp}) gives for $\mathcal{D}_{\perp,\uptau_z}$
\begin{equation}
\mathcal{D}_{\perp,\uptau_z}=\frac{1}{2\pi\nu\tau^2}\times\frac{1}{D_0q^2+\omega-i\uptau_z\Delta_v+\Delta_\perp}\label{eqn:app_dperptau}
\end{equation}
where $\Delta_\perp=\tau_\parallel/\tau\tau_\perp$. 





\section{Diffusion corrections}\label{app:corrections}
%
The expressions for $\delta D$ and $\delta z$ extracted from the diagrams in Fig.\;\ref{fig:sigma_skeleton} for the $\mathcal{D}_+$ propagator are given below.  
\begin{widetext}
\begin{subequations}
\begin{eqnarray}
\frac{\delta D}{D}&=&-\frac{4}{\nu}\iint Dq^2\left[\mathcal{D}_{+}^3(q,\omega)\left(\Gamma_{1+}-\Gamma_{2+}\right)
+\mathcal{D}_{-}^3(q,\omega)\left(\Gamma_{1-}-\Gamma_{2-}\right)
+\mathcal{D}_{\perp}^3(q,\omega)\left(\Gamma_{1\perp}-2\Gamma_{2\perp}\right)\right]\label{eqn:deltaD}\\ 
\delta z&=&-\frac{1}{\pi\nu}\int\frac{d^2q}{(2\pi)^2}\left[\mathcal{D}_{+}(q,0)\left(\Gamma_{1+}-\Gamma_{2+}\right)
+\mathcal{D}_{-}(q,0)\left(\Gamma_{1-}-\Gamma_{2-}\right)
+\mathcal{D}_{\perp}(q,0)\left(\Gamma_{1\perp}-2\Gamma_{2\perp}\right)\right]\label{eqn:deltaz}
\end{eqnarray}\label{eqn:sigma_skeleton}
\end{subequations}
\end{widetext}
(Note the factor of two in front of $\Gamma_{2\perp}$; for convenience the $\tau_z$ index is suppressed in the $\perp$ terms.)
%
%
The diffusion corrections to the amplitudes $\Gamma_{i,\alpha}$, where $i=1,2$ and $\alpha=\pm$, are detailed below. 
\begin{subequations}\label{eqn:gamma_skeleton}
\begin{eqnarray}
\delta \Gamma_{1+}&=& \frac{1}{4\pi \nu}\int\frac{d^2q}{(2\pi)^2}\left[\Gamma_{2+}\mathcal{D}_{+}
+\Gamma_{2-}\mathcal{D}_{-}+2\Gamma_{2\perp}\mathcal{D}_{\perp}\right]\nonumber\\
&&+\frac{1}{\nu}\iint\Gamma_{2+}\left[\Gamma_{2+}\mathcal{D}_{+}^2+\Gamma_{2-}\mathcal{D}_{-}^2+2\Gamma_{2\perp}\mathcal{D}_{\perp}^2\right]-\frac{1}{2}\left[\Gamma_{2+}^2\mathcal{D}_{+}^2+\Gamma_{2-}^2\mathcal{D}_{-}^2+2\Gamma_{2\perp}^2\mathcal{D}_{\perp}^2\right]\nonumber\\
 &&-\frac{1}{\nu}\iint\omega\Gamma_{2+}\left[\Gamma_{2+}^2\mathcal{D}_{+}^3+\Gamma_{2-}^2\mathcal{D}_{-}^3+2\Gamma_{2\perp}^2\mathcal{D}_{\perp}^3\right]-\omega\Gamma_{2+}^2\left[\Gamma_{2+}\mathcal{D}_{+}^3+\Gamma_{2-}\mathcal{D}_{-}^3+2\Gamma_{2\perp}\mathcal{D}_{\perp}^3\right]\nonumber\\
&&-\frac{1}{2\nu}\iint\omega^2\Gamma_{2+}^2\left[\Gamma_{2+}^2\mathcal{D}_{+}^4+\Gamma_{2-}^2\mathcal{D}_{-}^4+2\Gamma_{2\perp}^2\mathcal{D}_{\perp}^4\right]\\
\delta \Gamma_{2+}&=& \frac{1}{\pi \nu}\int\frac{d^2q}{(2\pi)^2}\left[\Gamma_{1+}\mathcal{D}_{+}
+\Gamma_{1-}\mathcal{D}_{-}+\Gamma_{1\perp}\mathcal{D}_{\perp}\right]\nonumber\\
&&+\frac{4}{\nu}\iint\Gamma_{2+}\left[\Gamma_{2+}\mathcal{D}_{+}^2+\Gamma_{2-}\mathcal{D}_{-}^2+2\Gamma_{2\perp}\mathcal{D}_{\perp}^2\right]-\frac{1}{2}\left[\Gamma_{2+}^2\mathcal{D}_{+}^2+\Gamma_{2-}^2\mathcal{D}_{-}^2+2\Gamma_{2\perp}^2\mathcal{D}_{\perp}^2\right]\nonumber\\
 &&-\frac{4}{\nu}\iint\omega\Gamma_{2+}\left[\Gamma_{2+}^2\mathcal{D}_{+}^3+\Gamma_{2-}^2\mathcal{D}_{-}^3+2\Gamma_{2\perp}^2\mathcal{D}_{\perp}^3\right]-\omega\Gamma_{2+}^2\left[\Gamma_{2+}\mathcal{D}_{+}^3+\Gamma_{2-}\mathcal{D}_{-}^3+2\Gamma_{2\perp}\mathcal{D}_{\perp}^3\right]\nonumber\\
&&-\frac{2}{\nu}\iint\omega^2\Gamma_{2+}^2\left[\Gamma_{2+}^2\mathcal{D}_{+}^4+\Gamma_{2-}^2\mathcal{D}_{-}^4+2\Gamma_{2\perp}^2\mathcal{D}_{\perp}^4\right]\\
\delta \Gamma_{1-}&=& \frac{1}{4\pi \nu}\int\frac{d^2q}{(2\pi)^2}\left[\Gamma_{2+}\mathcal{D}_{-}
+\Gamma_{2-}\mathcal{D}_{+}-2\Gamma_{2\perp}\mathcal{D}_{\perp}\right]\nonumber\\
&&+\frac{1}{\nu}\iint\Gamma_{2-}\left[\left(\Gamma_{2+}+\Gamma_{2-}\right)\mathcal{D}_+\mathcal{D}_--2\Gamma_{2\perp}\mathcal{D}_\perp^2\right]
-\frac{1}{2}\left[\Gamma_{2-}\Gamma_{2+}\left(\mathcal{D}_+^2+\mathcal{D}_-^2\right)-2\Gamma_{2\perp}^2\mathcal{D}_\perp^2\right]\nonumber\\
&&+\frac{4}{\nu}\iint\Gamma_{1-}\left[\left(\Gamma_{1+}-\Gamma_{2+}\right)\left(\mathcal{D}_+\mathcal{D}_--\mathcal{D}_+^2\right)+
\left(\Gamma_{1-}-\Gamma_{2-}\right)\left(\mathcal{D}_+\mathcal{D}_--\mathcal{D}_-^2\right)-2\left(\Gamma_{1\perp}-2\Gamma_{2\perp}\right)\mathcal{D}_\perp^2\right]\nonumber\\
 &&-\frac{1}{\nu}\iint\omega\Gamma_{2-}\left[\Gamma_{2-}
 \Gamma_{2+}\left(\mathcal{D}_-^2\mathcal{D}_++\mathcal{D}_-\mathcal{D}_+^2\right)
-2\Gamma_{2-}\Gamma_{2\perp}\mathcal{D}_\perp^3\right]-\omega\Gamma_{2-}^2\left[\Gamma_{2+}\mathcal{D}_{+}^2\mathcal{D}_-+\Gamma_{2-}\mathcal{D}_+\mathcal{D}_-^2-2\Gamma_{2\perp}\mathcal{D}_\perp^3\right]\nonumber\\
&&-\frac{1}{\nu}\iint\omega^2\Gamma_{2-}^2\left[\Gamma_{2-}\Gamma_{2+}\mathcal{D}_{+}^2\mathcal{D}_-^2-\Gamma_{2\perp}^2\mathcal{D}_{\perp}^4\right]\\
\delta \Gamma_{2-}&=& \frac{1}{\pi \nu}\int\frac{d^2q}{(2\pi)^2}\left[\Gamma_{1+}\mathcal{D}_{-}
+\Gamma_{1-}\mathcal{D}_{+}-\Gamma_{1\perp}\mathcal{D}_{\perp}\right]\nonumber\\
&&+\frac{4}{\nu}\iint\Gamma_{2-}\left[\Gamma_{2+}\mathcal{D}_{+}^2+\Gamma_{2-}\mathcal{D}_{-}^2+2\Gamma_{2\perp}\mathcal{D}_{\perp}^2\right]-\frac{1}{2}\left[\Gamma_{2-}\Gamma_{2+}\mathcal{D}_{+}^2+\Gamma_{2-}^2\mathcal{D}_{-}^2+2\Gamma_{2\perp}^2\mathcal{D}_{\perp}^2\right]\nonumber\\
&&+\frac{4}{\nu}\iint\Gamma_{2-}\left[\Gamma_{1+}\left(\mathcal{D}_{+}\mathcal{D}_- - \mathcal{D}_+^2\right)+\Gamma_{1-} \left(\mathcal{D}_{+}\mathcal{D}_- - \mathcal{D}_-^2\right) - 2\Gamma_{1\perp}\mathcal{D}_\perp^2\right]\nonumber\\
&&-\frac{4}{\nu}\iint\omega\Gamma_{2-}\left[\Gamma_{2+}^2\mathcal{D}_{+}^3+\Gamma_{2-}^2\mathcal{D}_{-}^3+2\Gamma_{2\perp}^2\mathcal{D}_{\perp}^3\right]-\omega\Gamma_{2-}\Gamma_{2+}\left[\Gamma_{2+}\mathcal{D}_{+}^3+\Gamma_{2-}\mathcal{D}_{-}^3+2\Gamma_{2\perp}\mathcal{D}_{\perp}^3\right]\nonumber\\
&&-\frac{2}{\nu}\iint\omega^2\Gamma_{2-}\Gamma_{2+}\left[\Gamma_{2+}^2\mathcal{D}_{+}^4+\Gamma_{2-}^2\mathcal{D}_{-}^4+2\Gamma_{2\perp}^2\mathcal{D}_{\perp}^4\right]
\end{eqnarray}

\end{subequations}


The terms above are ordered in correspondence with the diagrams appearing in Fig.\;\ref{fig:gamma_skeleton}. The  square brackets gather vertices that come together with the diffuson propagators. Note that the first term is unique in that it does not involve frequency integration. 
 (If these equations are calculated using perturbation theory, an additional wave-function renormalization term $\zeta$ appears\;\cite{pedagogical}. The renormalized amplitudes given below correspond to $\Gamma\zeta^2$, with $\delta \zeta = -\frac{2}{\nu}\iint \left(\Gamma_{1+}-\Gamma_{2+}\right)\mathcal{D}_+^2+\left(\Gamma_{1-}-\Gamma_{2-}\right)\mathcal{D}_-^2+\left(\Gamma_{1\perp}-2\Gamma_{2\perp}\right)\mathcal{D}_\perp^2$. It should be noted that the term $\zeta$ does not appear in the non-linear sigma model approach developed in Ref.\;\cite{sasha83}.)



\end{document}